\documentclass{aastex63}

 \DeclareUnicodeCharacter{0301}{*************************************}

\shorttitle{Detection of H$_2$ in the TWA 7 Debris Disk System}
\shortauthors{Flagg et al.}

\begin{document}

%% LaTeX will automatically break titles if they run longer than
%% one line. However, you may use \\ to force a line break if
%% you desire.

\title{Detection of H$_2$ in the TWA 7 System: A Probable Circumstellar Origin}

%% Use \author, \affil, and the \and command to format
%% author and affiliation information.
%% Note that \email has replaced the old \authoremail command
%% from AASTeX v4.0. You can use \email to mark an email address
%% anywhere in the paper, not just in the front matter.
%% As in the title, use \\ to force line breaks.

\author{Laura Flagg}
%\author{Laura Flagg\altaffilmark{1,2,}}
\affil{Department of Physics and Astronomy, Rice University, 6100 Main St. MS-108, Houston, TX 77005, USA}
\email{laura.flagg@rice.edu}

\author{Christopher M. Johns-Krull}
%\author{Laura Flagg\altaffilmark{1,2,}}
\affil{Department of Physics and Astronomy, Rice University, 6100 Main St. MS-108, Houston, TX 77005, USA}

\author{Kevin France}
\affil{Laboratory for Atmospheric and Space Physics, University of Colorado, 600 UCB, Boulder, CO 80309, USA}

\author{Gregory Herczeg}
\affil{Kavli Institute for Astronomy and Astrophysics, Peking University, Yi He Yuan Lu 5, Haidian Qu, Beijing 100871, China}

\author{Joan Najita}
\affil{National Optical Astronomy Observatory, 950 Cherry Avenue, Tucson, AZ. 85719, USA}

\author{John M. Carpenter}
\affil{Joint ALMA Observatory, Avenida Alonso de $\rm{\acute{C}}$ordova 3170, Santiago, Chile}

\author{Scott J. Kenyon}
\affil{Smithsonian Astrophysical Observatory, 60 Garden Street, Cambridge, MA 02138 USA}

%\email{aastex-help@aas.org}
%
%\and
%
%\author{R. J. Hanisch}
%\affil{Space Telescope Science Institute, Baltimore, MD 21218}
%
%
%\altaffiltext{1}{Visiting Astronomer, Cerro Tololo Inter-American Observatory.
%CTIO is operated by AURA, Inc.\ under contract to the National Science
%Foundation.}
%\altaffiltext{2}{Society of Fellows, Harvard University.}

\begin{abstract}
\noindent Using HST-COS FUV spectra, we have discovered warm molecular hydrogen in the TWA 7 system.  TWA 7, a $\sim$9 Myr old M2.5 star, has a cold debris disk and has previously shown no signs of accretion.   Molecular hydrogen is expected to be extremely rare in a debris disk.  While molecular hydrogen can be produced in star spots or the lower chromospheres of cool stars such as TWA 7, fluxes from progressions that get pumped by the wings of Ly$\alpha$ indicate  that this  molecular hydrogen could be circumstellar and thus that TWA 7 is accreting at very low levels and may retain a reservoir of gas in the near circumstellar environment.
\end{abstract}

%% Keywords should appear after the \end{abstract} command. The uncommented
%% example has been keyed in ApJ style. See the instructions to authors
%% for the journal to which you are submitting your paper to determine
%% what keyword punctuation is appropriate.


\section{Introduction}
Planet formation and evolution are heavily dependent on the circumstellar environment.  The circumstellar material can dictate formation, composition, orbital parameters, and migration of planets.  While much has been learned in recent years about circumstellar disks, especially with ALMA \citep{AndrewsDiskSubstructuresHigh2018}, the evolution from protoplanetary disk to debris disk is not well understood.  This time period is crucial for the final stages of the growth of terrestrial planets and the early evolution of their atmospheres \citep[e.g.][]{KenyonTerrestrialPlanetFormation2006, OlsonNebularatmospheremagma2019}.  Additionally,  ALMA typically images only the outer regions of disks; it is also of interest to understand the inner few AU of a system where many exoplanets are found and where the systems' habitable zones are.

Protoplanetary disks consist of gas, dust, and eventually planetesimals.  All three components play a crucial role in the formation and evolution of planets. Measurements of micron-sized dust are relatively easy, as dust can produce detectable amounts of infrared (IR) emission, even through the debris disk stage.  However, the primordial gas, which is mostly molecular hydrogen, is assumed to be 99\% of the protoplanetary disk mass \citep[see review by][]{WilliamsProtoplanetaryDisksTheir2011} and controls much of the disk dynamics, such as altering orbits of planetesimals and planets \citep[e.g.,]{WeidenschillingAerodynamicssolidbodies1977,  GoldreichEccentricityEvolutionPlanets2003,YoudinStreamingInstabilitiesProtoplanetary2005, BaruteauPlanetDiskInteractionsEarly2014} and potentially producing rings and spirals \citep{LyraFormationsharpeccentric2013, GonzalezSelfinduceddusttraps2017}.  Lower gas fractions and optically thin gas are expected in debris disks \citep{WyattEvolutionDebrisDisks2008}, although the precise gas fraction is poorly constrained \citep{MatthewsObservationsModelingTheory2014} and possibly varies significantly between disks. But even small amounts of optically thin gas can still have a large effect on disk dynamics \citep[e.g.][]{TakeuchiDustMigrationMorphology2001,  LyraFormationsharpeccentric2013}.  Thus, to understand the evolution of the circumstellar environment, we must understand how the hydrogen evolves.

%  Dust is also used as a tracer for smaller planetesimals.

However, H$_2$ is notoriously hard to detect.  Its only allowed electric dipole transitions are in the ultraviolet (UV).  In most circumstellar environments, those transitions require excited H$_2$, which can occur in circumstellar disks with warm gas \citep{NomuraMolecularhydrogenemission2005, AdamkovicsFUVIrradiatedDisk2016}. Warm gas is common in protoplanetary disks, but is less likely to be found in debris disks because of the generally large distance of the gas from the central star.    Chromospheric and transition region lines, such as Ly$\alpha$, pump the H$_2$ molecule from an excited level in the ground electronic state to the first (Lyman) or second (Werner) electronic levels.  Because of extremely high oscillator strengths, the excited molecule immediately decays back to the ground electronic level in a fluorescent cascade, emitting photons.  The set of emission lines produced by transitions from a single excited electronic state to the multiple allowed ground electronic states is called a progression.    Within a given progression,  H$_2$ line fluxes are proportional to their branching ratios \citep{WoodAnalysisH2Emission2002, HerczegFarUltravioletSpectraTW2004}. Because of this, far UV spectra are a powerful way to characterize the warm H$_2$ gas.  Emission from these lines is a probe of gas temperature.  But as all these transitions are in the UV, they require data from space-based observatories, thus limiting the number of observations currently available.  There are magnetic quadrapole transitions in the IR that have been detected in protoplanetary disks \citep[e.g][]{WeintraubDetectionQuiescentMolecular2000, BaryDetectionsRovibrationalH22003}; unfortunately, they are  weak and require much larger amounts of warm H$_2$ than debris disks typically have in order to detect them \citep[e.g.][]{BitnerTEXESSurveyH22008, Carmonasearchmidinfraredmolecular2008}.

%Debris disks also have less gas.%

To try to get around these issues, other molecules, most notably IR and millimeter transitions of HD and more commonly CO, have been used to trace the H$_2$ \citep[e.g.][]{TrapmanFarinfraredHDemission2017}.  However, neither is a perfect tracer, and both rely on an assumed ratio to H$_2$.  For example, disk mass estimates have often used the ISM CO/H$_2$ of $\sim$10$^{-4}$, consistent with the value found by \citet{FranceCOH2Abundance2014} based on CO and H$_2$ observations in the UV, but other recent studies have shown that CO appears depleted in protoplanetary disks \citep{FavreSignificantlyLowCO2013,  SchwarzUnlockingCODepletion2019, McClureCarbondepletionobserved2019}.  Furthermore, the difference in chemistry and masses between the molecular species mean that neither HD or especially CO trace H$_2$ perfectly \citep{MolyarovaGasMassTracers2017, AikawaMultiplePathsDeuterium2018}.

Molecular hydrogen emission has been detected in every protoplanetary and transition disk that have  far UV spectral observations \citep[e.g.][]{ValentiIUEAtlasPre2000, ArdilaObservationsTauriStars2002, HerczegOriginsFluorescentH22006, InglebyFarUltravioletH2Emission2009, FranceHubbleSpaceTelescope2012, YangFarultravioletAtlasLowresolution2012, France1600EmissionBump2017}. Debris disks are not defined by their gas content --- they are instead defined by secondary dust produced from planetesimal collisions, which observationally gets translated into a  fractional luminosity, $f=L_{disk}/L_*$ less than 10$^{-2}$ --- but all evidence indicates that compared with protoplanetary disks, they have a smaller gas-to-dust ratio and less gas in total \citep[e.g.][]{ChenDustGasPictoris2007}.  Other gas species, like CO,  have been detected in debris disks \citep[e.g.][]{RobergeDebrisDisksNearby2008, MoorMolecularGasYoung2011, DentMolecularGasClumps2014, HiguchiDetectionSubmillimeterwaveEmission2017}, but the only previous potential detection of H$_2$ in what is clearly a debris disk is from AU Mic \citep{FranceLowMassH2Component2007}.  This is not unexpected.   While  comets in our own Solar System produce CO, they do not produce H$_2$ \citep{MummaChemicalCompositionComets2011}. Thus, it is likely that secondary H$_2$ is not produced in the same manner as secondary CO.  In several cases, there are arguments for the reclassification of systems based on the discovery of H$_2$, such as RECX 11 \citep{InglebyEvolutionXrayFarultraviolet2011},  HD 98800 B (TWA 4 B) \citep{YangFarultravioletAtlasLowresolution2012, RibasLonglivedProtoplanetaryDisks2018}, and potentially DoAr 21 \citep{BaryDetectionsRovibrationalH22003, JensenNoTransitionDisk2009}. But exactly when and on what timescale the H$_2$ dissipates is not known.

%Debris disks are defined by secondary dust produced from planetesimal collisions instead of primordial dust.  Observationally, this has been translated into a fractional luminosity, $f=L_{disk}/L_*$ less than 10$^{-2}$.  While neither version of the definition relies on gas, 

Since even small amounts of H$_2$ gas can have a significant impact on planetary systems at ages $\sim$10 Myr, we have begun a program to examine UV spectra of young stars that show no evidence of near-infrared (NIR) excess.  One specific way that gas can impact a system is by limiting the IR flux from dust produced by planetesimal collisions.  \citet{KenyonRockyPlanetFormation2016} show that there is a discrepancy between the incidence rate of dust expected to be produced by terrestrial planet formation (2 to 3\% of young systems) and the incidence rate of close-in terrestrial planets (20\% of mature systems).  Gas, however, could sweep away that dust via gas drag, making it harder to detect. Thus, it is critical to understand the evolution of H$_2$ in the terrestrial planet forming regions.

%In order to detect any potential warm gas, we analyzed  Hubble Space Telescope (HST) UV spectra of 5-10 Myr stars with no near-Infrared (IR) excess to look for H$_2$.

\section{Target and Observations}

  TWA 7 is an M dwarf that is part of the $\sim$7-10 Myr TW Hya Association \citep{WebbDiscoverySevenTauri1999}. Recent spectral classifications assign an M2 or M3 spectral type \citep{ManaraXshooterspectroscopyyoung2013, HerczegOpticalSpectroscopicStudy2014}; we adopt M2.5.  The star is surrounded by a debris disk that was first detected due to its IR excess at 24 and 70 $\mu$m by \citet{LowExploringTerrestrialPlanet2005} with the Spitzer Space Telescope.  However, the lack of near IR excess \citep{WeinbergerSearchWarmCircumstellar2004} and typical accretion signatures \citep{JayawardhanaAccretionDisksYoung2006} strongly imply that it is a ``cool'' debris disk, making it one of the few known M stars with a debris disk \citep{TheissenWarmDustCool2014}.  The dust in the disk has since been detected in the FIR at 450 and 850 $\mu$m using the James Clerk Maxwell Telescope  \citep{MatthewsMassTemperatureTWA2007} and at 70, 160, 250, and 350 $\mu$m using the Herschel Space Observatory  \citep{CiezaHerschelDIGITSurvey2013}.  No [O I] was detected by Herschel  at 63 $\mu$m \citep{Riviere-MarichalarGasdustTW2013}, but CO has been recently  detected using ALMA in the J=3-2 transition \citep{MatraUbiquityStellarLuminosity2019}. The disk has been imaged in the IR with SPHERE showing spiral arms near 25 AU \citep{OlofssonResolvingfaintstructures2018}. \citet{YangFarultravioletAtlasLowresolution2012} and \citet{FranceHubbleSpaceTelescope2012} both failed to detect H$_2$ around TWA 7 in UV spectra.  \citet{YangFarultravioletAtlasLowresolution2012} used a less sensitive prism spectrum;  \citet{FranceHubbleSpaceTelescope2012} looked at 12 H$_2$ features separately, as opposed to detecting the combined H$_2$ emission from many features as we do in this paper. (See Section \ref{cc_sec}.)

 %Given that it has not yet been measured in many systems, it is hard to determine what role H$_2$ plays in debris disks.  However, w
 
 We can put some constraints on the expected H$_2$ based on dust and CO measurements.  For TWA 7, the total dust mass in the disk, M$_d$, is 2$\times$10$^{-2}$ M$_\oplus$ \citep{BayoSubmillimetrenoncontaminateddetection2019}, while the mass of CO in the disk, M$_{CO}$, is 0.8-80$\times$10$^{-6}$ M$_\oplus$ \citep{MatraUbiquityStellarLuminosity2019}.  Based on these estimates, if TWA 7 has an ISM value for the CO/H$_2$ ratio of $\sim$10$^{-4}$ then we can expect M$_{H_2}$ to be on the order of  M$_d$.  If it has a lower CO/H$_2$ of $\sim$10$^{-6}$, as TW Hya has \citep{FavreSignificantlyLowCO2013}, we can expect M$_{H_2}$  to be 100$\times$ larger than M$_d$, consistent with the ISM gas-to-dust ratio \citep{SpitzerPhysicalprocessesinterstellar1978}.  Models that have explored gas-to-dust ratios between 0.01 and 100 indicate that gas can significantly influence the disk dynamics  \citep{YoudinStreamingInstabilitiesProtoplanetary2005, LyraFormationsharpeccentric2013, GonzalezSelfinduceddusttraps2017}, so in either case, H$_2$ could  play an important if not dominant role in TWA 7's disk dynamics. Given the presence of this distant reservoir of gas, we explore here the possibility that an (as yet unseen) reservoir of gas is also present at smaller disk radii, in the terrestrial planet region of the disk.% \textbf{Based on \citet{BrudererSurvivalmoleculargas2013}, we expect the H$_2$ we could detect --- the ``warm H$_2$'' at around 1500 K --- to be about 10$^{-6}$ of the total H$_2$ for the entire disk.}

\begin{figure}
\centering
\includegraphics[width=6.4in]{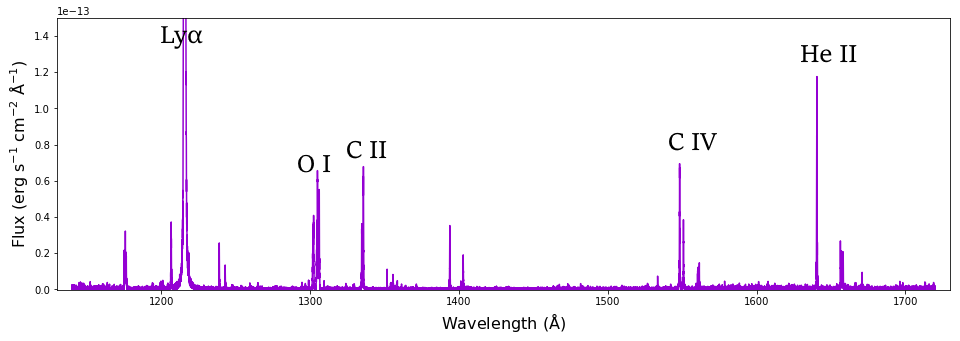}
\caption{The spectrum of TWA 7 used for analysis with the most prominent stellar features labeled.  Note that the Ly$\alpha$ profile is largely  geocoronal airglow emission and was thus not used. \label{spectrum}}
\end{figure}

%20110505
For this work, we use archival HST-Cosmic Origins Spectrograph  (COS) observations of TWA 7 from May 2011 (PID 11616, PI: G. Herczeg).  The data were acquired with the far UV medium resolution modes of COS: G130M and G160M.  These spectra have a spatial resolution of 1'' and a wavelength uncertainty $\sim$15 km/s \citep{cosmic2020}. The observations are at a range of central wavelengths that allow us to get a contiguous spectrum that spans from 1133 to 1795 \AA\ (Figure \ref{spectrum}).  In addition to TWA 7, we also analyze spectra of classical T Tauri stars (CTTS) and main sequence M dwarf stars for comparison purposes (Table \ref{compstars}) taken between December 2009 and  August 2015.  The CTTS were chosen from the stars analyzed by \citet{FranceHubbleSpaceTelescope2012} that had extinction values measured by both \citet{HerczegOpticalSpectroscopicStudy2014} and \citet{FurlanSpitzerInfraredSpectrograph2011}.  The main sequence M dwarfs were from \citet{KruczekH2FluorescenceDwarf2017}, chosen because they had H$_2$ detected from the stellar photosphere and COS spectra that covered a comparable wavelength range. One of the six M dwarfs --- GJ 581 --- has a cold, faint debris disk \citep{LestradeDEBRISdiskplanet2012}, but it is much older (2-8 Gyr) and less active \citep{SchoferCARMENESsearchexoplanets2019} than TWA 7 or the CTTS.  Its disk is also significantly less luminous than that of TWA 7 \citep{ChoquetFirstimagesdebris2016}. The remaining five M dwarfs have no detected disks.  All spectra were observed with COS in a similar manner.  Spectra were reduced by the CALCOS pipeline.  Multiple observations were then co-added into one spectrum as described by \citet{DanforthEmpiricallyEstimatedFarUV2016}.  The TWA 7 spectrum we analyzed is plotted in Figure \ref{spectrum}.

We also used archival HST-STIS spectra of TW Hya, reduced with the STIS pipeline. For each observation, we combined the orders to create a single spectrum. We then co-added the observations in a similar manner to the way we co-added the observations from COS.  

%, weak-lined T Tauri stars (WTTS)
 \begin{deluxetable}{llrrrl}
 \tabletypesize{\small}
 \tablewidth{0pt}
 \tablecaption{Stellar Properties}
 \tablehead{
 \colhead{Object} & \colhead{PID/PI} & \colhead{Distance} & \colhead{RV} &  \colhead{A$_V^a$} &  \colhead{A$_V^b$} 
\vspace{-5pt}
 \\
  \colhead{} & \colhead{} & \colhead{(pc)} & \colhead{(km s$^{-1}$)} &  \colhead{(mag)} &  \colhead{(mag)} 
  }
  \startdata
  \textbf{TWA 7}	&	\textbf{11616/Herczeg}	&	\textbf{34.0}	&	\textbf{11.4}	&	\nodata	&	\textbf{0.00}$^c$	\\
  %\hline
  \cutinhead{Classical T Tauri Stars}
AA Tau	&	11616/Herczeg	&	136.7	&	17.0	&	1.9	&	0.40	\\
BP Tau	&	12036/Green	&	128.6	&	15.2	&	1.0	&	0.45	\\
DE Tau	&	11616/Herczeg	&	126.9	&	15.4	&	0.9	&	0.35	\\
DM Tau	&	11616/Herczeg	&	144.5	&	18.6	&	0.0	&	0.10	\\
DR Tau	&	11616/Herczeg	&	194.6	&	21.1	&	1.4	&	0.45	\\
GM Aur	&	11616/Herczeg	&	159.0	&	15.2	&	0.6	&	0.30	\\
HN Tau	&	11616/Herczeg	&	136.1	&	4.6	&	1.0	&	1.15	\\
LkCa 15	&	11616/Herczeg	&	158.2	&	17.7	&	1.0	&	0.30	\\
SU Aur	&	11616/Herczeg	&	157.7	&	14.3	&	0.9	&	0.65	\\
UX Tau	&	11616/Herczeg	&	139.4	&	15.5	&	0.5	&	0.00$^c$	\\
%  \cutinhead{Weak-lined T Tauri Stars}
%HBC 427	&	11616	&	148.1	&	\nodata	&	\nodata	&	\nodata	\\
%LkCa 4	&	11616	&	129.4	&	\nodata	&	\nodata	&	\nodata	\\
%LkCa 19	&	11616	&	159.7	&	\cdots	&	\cdots	&	\nodata	\\
%RECX 1	&	11616	&	99.9	&	\nodata	&	\nodata	&	\nodata	\\
%TWA 13	&	12361	&	59.8	&	\nodata	&	\nodata	&	\nodata	\\
%TWA 13	&	12361	&	59.9	&	\nodata	&	\nodata	&	\nodata	\\
\cutinhead{Main Sequence M Stars with H$_2$}
GJ 176	&	13650/France	&	9.5	&	26.2	&	\nodata	&	\nodata	\\
GJ 832	&	12464/France	&	5.0	&	13.2	&	\nodata	&	\nodata	\\
GJ 667 C	&	13650/France	&	7.2	&	6.4	&	\nodata	&	\nodata	\\
GJ 436	&	13650/France	&	9.8	&	9.6	&	\nodata	&	\nodata	\\
GJ 581	&	13650/France	&	6.3	&	-9.4	&	\nodata	&	\nodata	\\
GJ 876	&	12464/France	&	4.7	&	-1.6	&	\nodata	&	\nodata	\\
\cutinhead{STIS Spectra}
TW Hya	&	11608/Calvet	&	60.1	&	13.4	&	\nodata	& 0.00	
\enddata
\tablecomments{Properties of stars analyzed in this paper.  \\
$^a$ A$_V$ from \citet{FurlanSpitzerInfraredSpectrograph2011}\\
$^b$ A$_V$ from \citet{HerczegOpticalSpectroscopicStudy2014} with uncertainties of 0.15 mag.\\
$^c$ The measured value was negative.  Since this is unphysical, we adapted an extinction of 0.0 mag. \\
Distances are from \citep{Bailer-JonesEstimatingDistanceParallaxes2018} based on Gaia DR2 \citet{CollaborationGaiaDataRelease2018}.  RVs of the T Tauri stars are from \citet{NguyenCloseCompanionsYoung2012}, from Gaia DR2 for the M stars, and from \citet{TorresSearchassociationscontaining2006} for TW Hya and TWA 7.
Based on extinction measurements from stars in the Local Bubble \citep{LeroyPolarimetricInvestigationInterstellar1993}, we assume these Main Sequence M stars have no extinction. 
\label{compstars}}
 \end{deluxetable}
 
 % nweeded: object name, category, RV, A_v, distance
 % possible: PID, UTdate, coords, H2 (will be in paper, unsure if it needs to be in table), mass acretion, inclination, vsini, age??

\section{Analysis and Results}
\subsection{Methods for Cross-Correlation}\label{cc_sec}
\begin{figure}
\centering
\includegraphics[width=6.4in]{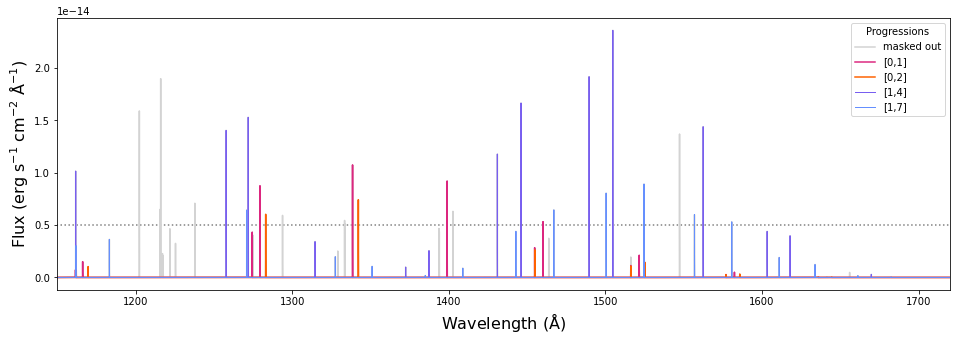}
\caption{Templates of the most prominent progressions used for cross-correlation.  A cutoff for a minimum line strength of 5$\times$10$^-{15}$ erg s$^{-1}$ cm$^{-2}$ \AA$^{-1}$ is also shown. \label{template}}
\end{figure}

\begin{figure}
\centering
\includegraphics[width=6.4in]{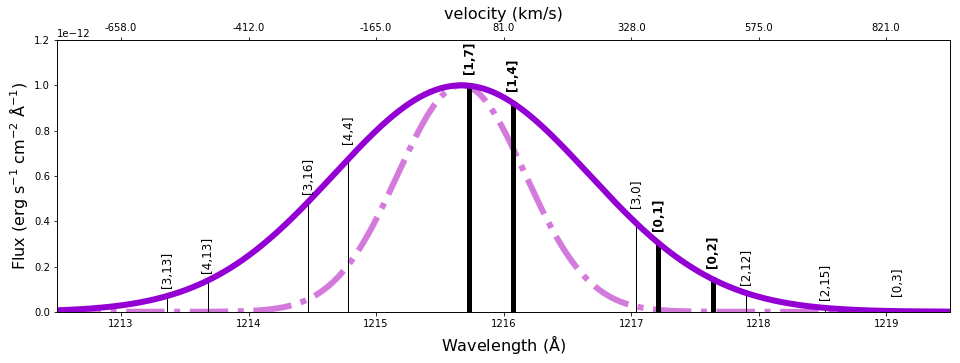}
\caption{Schematic Ly$\alpha$ profiles.  The solid purple line corresponds to a broader profile, like seen with CTTS, while the dashed pink line is indicative of Main Sequence stars with a more narrow profile.   The pumping wavelengths of some prominent are marked. The strongest progressions have not only substantial Ly$\alpha$ flux but also larger oscillator strengths and smaller lower state energies (Table \ref{pros}).   \label{cartoon}}
\end{figure}

In protoplanetary disks and nearby M dwarfs, the strength of the H$_2$ lines make them clearly detectable above the noise; however, this is not the case for systems with smaller amounts of H$_2$ flux. Instead, we take advantage of the many weak H$_2$ lines in the system and use a cross-correlation function (CCF), a technique that has been used previously with IR data to study gas in protoplanetary disks \citep{HartmannFurtherevidencedisk1987}.  The CCF allows us to combine the signal from multiple lines into one signal by calculating how well the spectrum correlates with that of a model template \citep{Tonrysurveygalaxyredshifts1979}.   Our full template was created using the procedure from \citet{McJunkinEmpiricallyEstimatedFarUV2016} for a temperature of 2500 K and a column density of log$N$(H$_2$)=19 cm$^2$. As temperature and density have little effect on the relative strengths of these lines in  protoplanetary disks \citep{FranceHubbleSpaceTelescope2012}, we used the same template for all the stars. Ly$\alpha$ also has a significant impact on H$_2$ line strengths, but its profile is contaminated from self-absorption, ISM absorption, and geocoronal airglow, and thus cannot be used for all of our targets. As a result, we do not consider the shape of the Ly$\alpha$ profile when defining our template. We use a single FWHM of 0.047 \AA\ for the lines in the template, chosen solely because it is the width that maximizes the CCF for TWA 7.  Templates for individual progressions were created by picking the lines from the full template based on \citet{AbgrallTableLymanBand1993}  (Figure \ref{template}). Although there are many H$_2$ lines, focusing only on the strongest lines gives the clearest signal. We chose the minimum H$_2$ line strength in the template that maximized the CCF detection for TWA 7 for each progression.  The minimum line strength is dependent on how strong the lines in that progression are: progressions with weaker fluxes require smaller minimum line strengths.   While we analyzed 12 different progressions (Table \ref{pros}), chosen because all 12 were detected by \citet{FranceHubbleSpaceTelescope2012} in protoplanetary disks, our analysis focused on the progressions that typically produce the most H$_2$ flux in CTTS --- [1,4], [1,7], [0,1], and [0,2].  Each of these progressions is excited by Ly$\alpha$  and can decay to multiple lower states, resulting in a set of H$_2$ emission lines throughout the UV.  The total summed flux in an individual progression is a function of having enough Ly$\alpha$ photons to pump the H$_2$ molecule to the excited state (Figure \ref{cartoon}), the filling factor of H$_2$ around the Ly$\alpha$, the column density in the excited rovibrational level of the X electronic state, and the oscillator strength of the the pump transition \citep{HerczegOriginsFluorescentH22006}.

%, assuming that the template line width that matches TWA 7's true line width best will produce the highest CCF peak

 \begin{deluxetable}{lcrcclc}
 \tabletypesize{\small}
 \tablewidth{0pt}
 \tablecaption{Progressions Analyzed}
 \tablehead{
 \colhead{[\hbox{$v^\prime$},\hbox{$J^\prime$}]} & \colhead{$\lambda_{pump}$} & \colhead{velocity} &  \colhead{TW Hya H$_2$ Flux}  & \colhead{oscillator strength} & \colhead{[\hbox{$v^{\prime\prime}$},\hbox{$J^{\prime\prime}$}]} & \colhead{E$^{\prime\prime}$} 
\vspace{-5pt}
 \\
  \colhead{} & \colhead{(\AA)} &  \colhead{(km s$^{-1}$)}  & \colhead{(10$^{-15}$ erg cm$^{-2}$ s$^{-1}$)} &  \colhead{($\times$ 10$^{-3}$)} &  \colhead{} &  \colhead{(eV)} 
  }
  \startdata
$[3,13]$	&	1213.356	&	-571	&\ \	4.7	&	20.6 & [1,14] & 1.79	\\
$[4,13]$	&	1213.677	&	-491	&\ \	2.4	& \quad	9.33 & [2,12] & 1.93	\\
$[3,16]$	&	1214.465	&	-297	&	14.9	&	23.6 & [1,15] & 1.94	\\
$[4,4]$	&	1214.781	&	-219	&\ \	8.9	&	\quad 9.90 & [3,5] & 1.65	\\
$[1,7]$	&	1215.726	&	14	&	16.2	&	34.8 & [2,6] & 1.27	\\
$[1,4]$	&	1216.070	&	99	&	36.0	&	28.9 & [2,5] & 1.20	\\
$[3,0]$	&	1217.038	&	338	& \  \	3.5	&	 \quad 1.28	 &[3,1] & 1.50 \\
$[0,1]$	&	1217.205	&	379	&	37.9	&	44.0 & [2,0] & 1.00	\\
$[0,2]$	&	1217.643	&	487	&	33.4	&	28.9 & [2,1] & 1.02	\\
$[2,12]$	&	1217.904	&	551	&	18.4	&	19.2	& [1,13] &  1.64\\
$[2,15]$	&	1218.521	&	704	& \ \ 	3.1	&	18.0 & [1,14] & 1.79	\\
$[0,3]$	&	1219.089	&	844	&	\ \ 2.1	&	25.5 &[2,2] & 1.04	\\
\enddata
\tablecomments{Velocity is from Ly$\alpha$ center.  TW Hya H$_2$ Flux as measured by \citet{HerczegOriginsFluorescentH22006}. Oscillator strengths of the pumping transitions calculated by \citet{HerczegOriginsFluorescentH22006} based on \citet{AbgrallTableLymanBand1993}.  \\
$[$v$^{\prime\prime}$,J$^{\prime\prime}$] and E$^{\prime\prime}$ are the lower level in the electronic ground state for the pumping transition and the corresponding energy for that state.\\
 Each of these progressions is pumped by Ly$\alpha$ flux and can decay to multiple lower states, resulting in a set of H$_2$ emission lines throughout the UV. 
\label{pros}}
 \end{deluxetable}

Since we want to be sure that we are only cross-correlating continuum and H$_2$ emission (plus the associated noise)  and not emission from hot gas lines from the chromosphere or transition region, we masked out FUV lines commonly seen in lower mass stars from \citet{HerczegFarUltravioletSpectrumTW2002}, \citet{BrandtDor94Hubble2001}, and \citet{AyresFarUltravioletUpsDowns2015}.  As these lines have different widths in different stars depending on numerous properties, we erred on the side of masking the wavelength regions covered by the broadest of these features to minimize the chance of a false positive from a line that was not H$_2$.

Cross-correlating the entire spectrum with the entire masked template returns a  tentative detection.  However, this involves cross-correlating a significant amount of noise which can weaken the detection.  Therefore we created segments of spectrum for each H$_2$ feature of $\sim$1 \AA\ ($\sim$200 km/s) wide centered around the wavelengths of expected H$_2$ lines, which is wide enough to get the entire line profile without adding too much continuum flux or noise.  (We cannot be certain as to whether the photons detected outside of lines are from the star itself, as M dwarfs have very little stellar continuum flux in these regions, so we will refer to the region outside of lines as ``continuum/noise.'') To calculate the final CCF, we explored two procedures to verify any findings.  With both methods, if the flux for every line is emitted at a similar relative velocity (within $\sim$10 km s$^{-1}$), the CCF's signal will grow stronger.  In the first method, we created one long spectrum by putting all the individual segments end-to-end.  We do the same for the corresponding template segments. We then cross-correlated this pieced together spectrum with the same regions from the template.  In the second method, we cross-correlated each segment of spectrum individually with its corresponding template segment and added the cross-correlation functions.  Because of this, we chose to use a CCF that has not been normalized for length, which is usually the last step of calculating the CCF. Unnormalized CCFs work equally well for both of our methods; normalized CCFs of different lengths cannot be added linearly, because longer CCFs should be weighted more.% We also define a minimum line flux for the H$_2$ line that maximizes the S/N of the CC to avoid lines that are so weak that we are just analyzing noise.

\subsection{\texorpdfstring{H$_2$ Detection and Verification}{H2 Detection and Verification}}
  \begin{figure}
\centering
\includegraphics[width=6.4in]{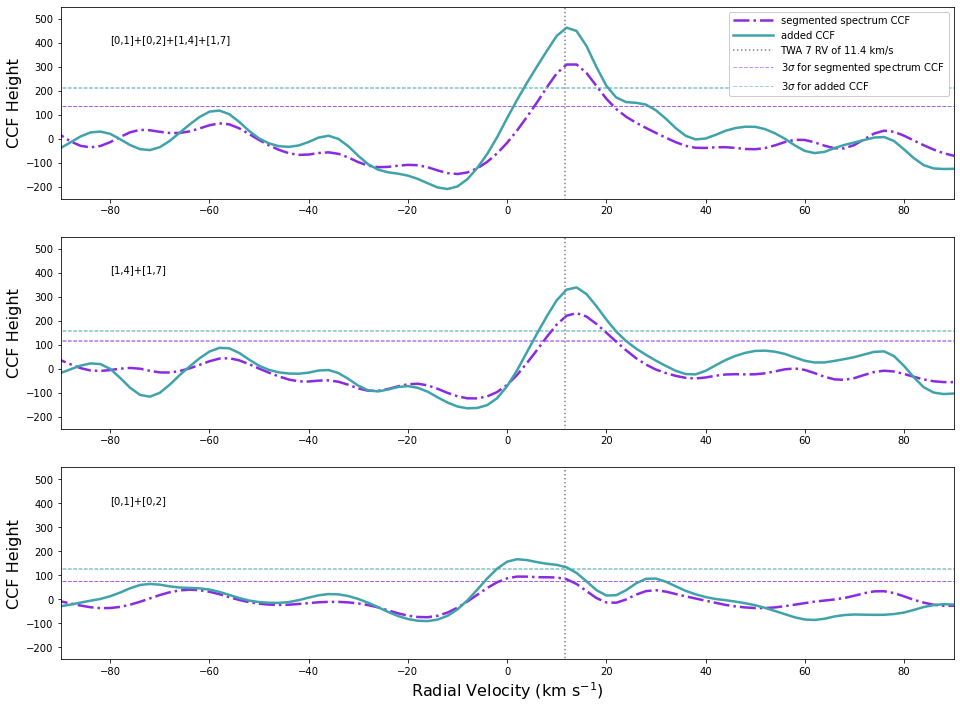}
\caption{Cross-correlation functions of TWA 7 for the most prominent H$_2$ progressions. \label{ccf}}
\end{figure}
We detect peaks near the stellar radial velocity in the CCFs of the spectra  of TWA 7 (Figure \ref{ccf}) using a minimum H$_2$ template line strength cutoff of 5$\times$10$^{-15}$ erg s$^{-1}$ cm$^{-2}$ \AA$^{-1}$.  The peaks are detected with both methods --- segmented spectrum and added CCF --- for calculating the CCF.  Although the peak is strongest when all of the most prominent progressions are included, we also see significant ($>$3$\sigma$) detections when some individual progressions are analyzed.  While the peak for [0,1]+[0,2] is slightly off-center from the systemic velocity of 11.4 km/s, we attribute this to the uncertainties in the wavelength calibration that can lead to shifts in the resulting velocities by up to 15 km/s, as described in \citet{LinskyUltravioletSpectroscopyRapidly2012}.% with just the [0,1] and [0,2] progressions.

\begin{figure}
\centering
\includegraphics[width=3.4in]{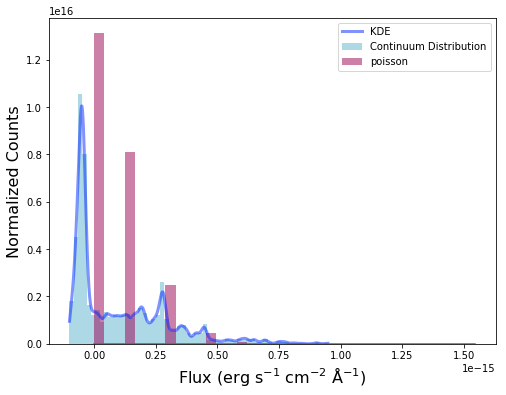}
\caption{Typical continuum/noise distributions, which we model in two ways: by creating a KDE of the distribution and by scaling a discrete Poisson distribution to the data.  Both were normalized so that the total area is equal to 1. \label{kde}}
\end{figure}

The strength of the cross-correlation function is dependent on the S/N in the  H$_2$ lines.  Since we are trying to measure the height and significance of the CCF peak, we need to understand the noise properties of the observed spectrum.  This is made more difficult, because there are so few FUV photons that reach us.  For example, \citet{LoydMUSCLESTreasurySurvey2016} looked for FUV continuum in our M dwarf sample, obtaining a significant detection for only 3 of 6 targets.  As a result, noise in our spectrum cannot be approximated to be Gaussian as it can when there are hundreds or thousands of counts. Typical continuum/noise regions in our TWA 7 spectrum have flux distributions that look approximately like that seen in  Figure \ref{kde} where we show a histogram of flux levels found in the continuum/noise of TWA 7. %Additionally, COS does not count photons --- it counts pulses which can be induced by photons.  

%To deal with this, we estimate the noise in three separate ways: with a Gaussian, with a scaled Poisson, and using the actually distribution fit with a kernel density smoother....???.., as shown in Figure \ref{kde}.

There are several potential issues in modeling this noise.  The first is that there are undoubtedly unidentified lines that we do not mask, as possibly seen in the increase around 0.3$\times10^{-15}$ erg s$^{-1}$ cm$^{-2}$ \AA$^{-1}$ .  However, since other unidentified lines could possibly overlap with the H$_2$ lines, we choose not to remove this peak from the distribution.  Another issue is that because of the low flux level, when the detector background gets subtracted, we end up measuring ``negative'' flux in some wavelength bins.   To deal with this, we estimate the noise in two separate ways:  with a scaled Poisson distribution and using the actual distribution fit with a kernel density estimator (KDE) \citep{RosenblattRemarksNonparametricEstimates1956, ParzenEstimationProbabilityDensity1962}, as shown in Figure \ref{kde}.  The scaled Poisson was determined by calculating the skew of the distribution of continuum/noise, $\gamma_1$.  The mean of the Poisson distribution $\lambda$ is then $\gamma_1^{-2}$. We then convert from counts to  flux using a constant scaling factor determined by the mean of the distribution. The KDE was calculated with a Gaussian kernel using a bandwidth (equivalent to the sigma parameter) of 10$^{-17}$ erg s$^{-1}$ cm$^{-2}$ \AA$^{-1}$. We randomly draw our noise from these distributions.  These two noise models cover the range of possibilities of the underlying true noise, so a robust detection will only  be evident if it occurs using both noise models.

\begin{figure}
\centering
\includegraphics[width=6.4in]{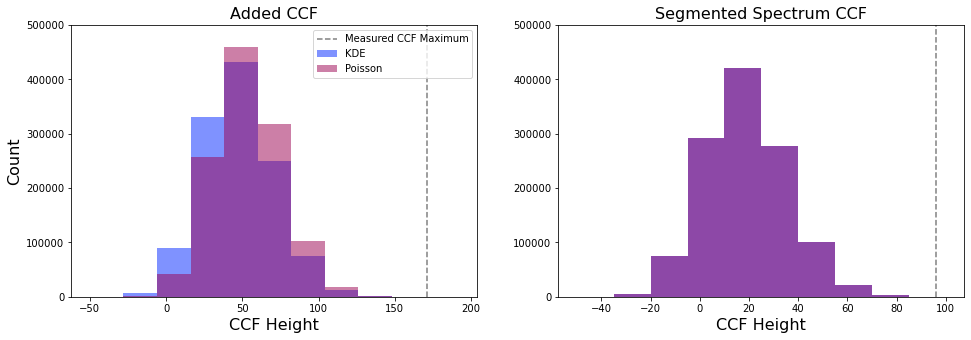}
\caption{Distribution of CCF heights for simulations of noise cross-correlated with  the [0,1] and [0,2] progressions of the H$_2$ template.   Based on 1,200,000 simulations for each set, we detect H$_2$ at a level of 4.5$\sigma$ for Poisson noise with the added CCFs, 4.6$\sigma$ for CDF sampled noise with the added CCFs, 3.9$\sigma$ for Poisson noise with the segmented spectrum CCF, and 4.0$\sigma$ for CDF sampled noise with the segmented spectrum CCF. \label{ccf_max_noise}}
\end{figure}

To determine the significance of the detection, we used our  noise models to create spectra containing only noise.  We then cross-correlated these noise spectra with the template in the same ways we did for the TWA 7 spectrum.  We then record the CCF maximum within 15 km s$^{-1}$ of the systemic velocity.  We chose this range, because this was the range we used to look for a detection, as COS has a velocity precision of 15 km s$^{-1}$.    This procedure was repeated multiple times (see Table \ref{pros_sig}) for each type of noise  to produce the distributions shown in Figure \ref{ccf_max_noise}.  We then compared the CCF maximums to TWA 7's CCF maximum within 15 km s$^{-1}$ of the systemic velocity.  The fraction of times the noise's CCF maximum was equal to or larger than TWA 7's CCF maximum is taken to be the probability of a false positive.  The significance ($\sigma$) values we report are the the equivalent probabilities for a Gaussian distribution.

%We match the height seen or higher with the real data for 1 in .... attempts for just.

We did an initial trial of 32,000 simulations with each method to see if we could detect each progression individually.  We obtain significant detections for [0,1], [0,2], [1,4], and [1,7], with significant being defined as $>$3$\sigma$ detections for all four methods; we also get a a marginal detection ($>$3$\sigma$ for some but not all methods) for [0,3] (Table \ref{pros_sig}).  We then investigated the detected progressions further.  Using a line strength cutoff of 5$\times$10$^{-15}$ erg s$^{-1}$ cm$^{-2}$ \AA$^{-1}$, as shown in Figure \ref{template}, we detect H$_2$ at a significance $>$5$\sigma$ for all of our noise models and CCF types for the combination of the [1,4], [1,7], [0,1], and [0,2] progressions based on 3,500,000 simulations of each.  For just the progressions on the wing --- [0,1] combined with [0,2]  --- we did 1,200,000 simulations for each measurement.  We detect H$_2$ at a level of 4.5$\sigma$ for Poisson noise with the added CCFs, 4.6$\sigma$ for KDE noise with the added CCFs, 3.9$\sigma$ for Poisson noise with the segmented spectrum CCF, and 4.0$\sigma$ for KDE sampled noise with the segmented spectrum CCF.   The segmented spectrum CCF produces similar distributions for both types of noise, as shown on the right of Figure \ref{ccf_max_noise}, because it is more robust to slight differences in noise models. % Since we considered the detection robust because it occured regardless of what type of CCF we used, the modeled noise has simulate a strong detection for \textit{both} the  added CCFs and the segmented spectrum CCF.  Therefore, we feel the 4.5$\sigma$ and 4.6$\sigma$ values are statistically more accurate. However, we cannot discriminate between the types of noise, so we chose to be conservative, and use the less significant number.  Thus we have adapted a significance of  4.5$\sigma$ for [0,1]+[0,2]. 

 \begin{deluxetable}{lrrrrrrc}
 \tabletypesize{\small}
 \tablewidth{0pt}
 \tablecaption{Detection Significance of H$_2$ in Progressions}
 \tablehead{
 \colhead{} & \multicolumn{2}{c}{\underline{Added CCF}} &  \multicolumn{2}{c}{\underline{Segmented Spectrum CCF}} & \colhead{} & \colhead{Minimum line strength} & \colhead{Lines}
\vspace{-5pt}
 \\
  \colhead{Progression} & \colhead{KDE} &  \colhead{Poisson}  & \colhead{  \hspace{7pt} KDE} &  \colhead{ \hspace{17pt}    Poisson}  &  \colhead{Simulations} &  \colhead{erg s$^{-1}$ cm$^{-2}$ \AA$^{-1}$}
  &\colhead{ Included}} 
  \startdata
$[3,13]$	&	2.2	&	2.2	& \hspace{2pt} 	1.6	&	1.5 \quad \hspace{2pt} & 32000 & 0.1$\times$10$^{-15}$ & 9	\\
$[4,13]$	&	0.7	&	0.7	& \hspace{2pt} 	0.8	&	0.8	 \quad \hspace{2pt}  & 32000	& 1.7$\times$10$^{-15}$ & 2 \\
$[3,16]$	&	2.4	&	2.4	& \hspace{2pt} 	1.8	&	1.8	 \quad \hspace{2pt}  & 32000 & 1.2$\times$10$^{-15}$ & 9	\\
$[4,4]$	&	1.2	&	1.2	& \hspace{2pt} 	2.0	&	2.0 \quad  \hspace{2pt}  & 32000	& 1.3$\times$10$^{-15}$ & 6	\\
$[1,7]$	&	$>$4.0	&	$>$4.0	& \hspace{2pt} 	$>$4.0	&	$>$4.0 \quad    \hspace{2pt} & 32000	& 7.5$\times$10$^{-15}$ & 2	\\
$[1,4]$	&	$>$4.0	&	$>$4.0	& \hspace{2pt} 	$>$4.0	&	$>$4.0 \quad  \hspace{2pt}  & 32000	& 3.0$\times$10$^{-15}$	& 12 \\
$[3,0]$	&	0.7	&	0.7	& \hspace{2pt} 	0.0	&	0.0	 \quad   \hspace{2pt}  & 32000	& 3.5$\times$10$^{-15}$ & 2 \\
$[0,1]$	&	3.5	&	3.4	& \hspace{2pt} 	3.2	&	3.1	  \quad  \hspace{2pt}  & 32000	& 9.0$\times$10$^{-15}$ & 2\\
$[0,2]$	&	$>$4.0	&	$>$4.0	& \hspace{2pt} 	3.3	&	3.3	  \quad  \hspace{2pt}  & 32000 & 5.0$\times$10$^{-15}$	&2 \\
$[2,12]$	&	2.2	&	2.3	& \hspace{2pt} 	2.6	&	2.6  \quad  \hspace{2pt}  & 32000	& 2.0$\times$10$^{-15}$ &2	\\
$[2,15]$	&	2.8	&	2.8	& \hspace{2pt} 	2.7	&	2.7  \quad  \hspace{2pt} 	 & 32000 & 3.0$\times$10$^{-15}$	& 4 \\
$[0,3]$	&	3.3	&	3.1	& \hspace{2pt} 	2.9	&	2.8  \quad   \hspace{2pt} & 32000	& 3.0$\times$10$^{-15}$ & 5	\\
\hline
$[0,1]+[0,2]$	&	4.6	&	4.5	& \hspace{2pt} 	4.0	&	3.9   \quad  \hspace{2pt} & 1200000 & 5.0$\times$10$^{-15}$ & 6	\\
$[1,4]+[1,7]+[0,1]+[0,2]$	&	$>$5.0	&	$>$5.0	& \hspace{2pt} 	$>$5.0	&	$>$5.0  \quad   \hspace{2pt} & 3500000	& 5.0$\times$10$^{-15}$ & 20 \\
\enddata
\tablecomments{Significance of detection in each individual progression for our four different methods, as well as for the combination of [0,1] and [0,2] and the combination of [1,4], [1,7], [0,1], and [0,2]. 
\label{pros_sig}}
 \end{deluxetable}

 There are two unclassified background objects in the sky near  TWA 7. However, the two sources are not expected to be in the COS-PSA aperture, given the COS-PSA aperture size of 2.5" in diameter and the distance of the background sources to TWA 7, at 2.5" \citep{Neuhauserpossibilitygroundbaseddirect2000} and 6" \citep{BayoSubmillimetrenoncontaminateddetection2019},  given we confirmed that TWA 7 was centered on the aperture.  Furthermore, as we required that the CCF peak at the radial velocity of the object fall within the error of the spectrograph,  we feel these background sources are an extremely unlikely source of the H$_2$.

\subsection{\texorpdfstring{Determining the Origin of the H$_2$}{Determining the Origin of the H2}}\label{origin}
Simply detecting H$_2$ does not indicate that the H$_2$ is circumstellar in origin, because  some M stars are known to show H$_2$ emission pumped by Ly$\alpha$ \citep{KruczekH2FluorescenceDwarf2017}.  Active M stars, like TWA 7 \citep{YangMagneticPropertiesYoung2008}, have strong chromospheric Ly$\alpha$ emission, which can pump H$_2$ in star spots or in their lower chromospheres.  Since TWA 7's debris disk is nearly face on at an inclination of 13$^\circ$ \citep{OlofssonResolvingfaintstructures2018}, and the resolution of COS is 15 km/s, we cannot use velocity information  to differentiate between circumstellar and stellar H$_2$.  Instead, we looked at the flux ratios between different progressions.  The [1,4] and [1,7] progression are both pumped by emission from the center of the Ly$\alpha$ line profile  (Figure \ref{cartoon}).  Other progressions  are pumped from the wings of the profile, so strong emission in these lines is only possible with a broader Ly$\alpha$ line indicative of active accretion.  The two most prominent examples are [0,1] and [0,2] which are pumped at velocities 379 and 487 km/s from line center.  These progressions should only be bright if the Ly$\alpha$ profile is especially wide, as shown in the purple profile in Figure \ref{cartoon}, but they will be much fainter if the star's Ly$\alpha$ profile is more narrow, similar to the pink curve.  Stars that are accreting have much broader  Ly$\alpha$ profiles than active main sequence stars \citep{SchindhelmLyaDominanceClassical2012, YoungbloodMUSCLESTreasurySurvey2016} and are therefore expected to produce more emission in the [0,1] and [0,2] progressions relative to the [1,4] and [1,7] progressions in comparison to non-accretors.

Using the segmented spectrum, we took the ratio of the CCF maximum of all the non-central progressions
 to the ratio of the CCF maximum of [1,4]+[1,7] for all the stars with previously detected H$_2$ in our sample. Dividing by the height of [1,4]+[1,7] for each system acts as a normalization factor to deal with spectra with different S/N or different line widths due to rotational broadening.   Since this ratio can be affected by extinction,  with lines at shorter wavelengths appearing fainter than they are intrinsically, we have to de-redden the spectrum first, which we do based on the extinction laws from \citet{CardelliRelationshipInfraredOptical1989}.   We examine three different sets of ratios: ratios with spectra uncorrected for extinction, ratios with spectra corrected by the extinction values found by \citet{FurlanSpitzerInfraredSpectrograph2011}, and ratios with spectra corrected by extinction values from \citet{HerczegOpticalSpectroscopicStudy2014}.  We assume the main sequence M stars and TWA 7 have no extinction based on extinction measurements from stars in the Local Bubble \citep{LeroyPolarimetricInvestigationInterstellar1993}. 
  %We took the next 4 most clear cut progressions --- [4,4], [3,16], [2,15], and [2,12] --- and performed a statistical analysis on just those as well as those with [0,1] and [0,2].   

\begin{figure}
\centering
\includegraphics[width=6.4in]{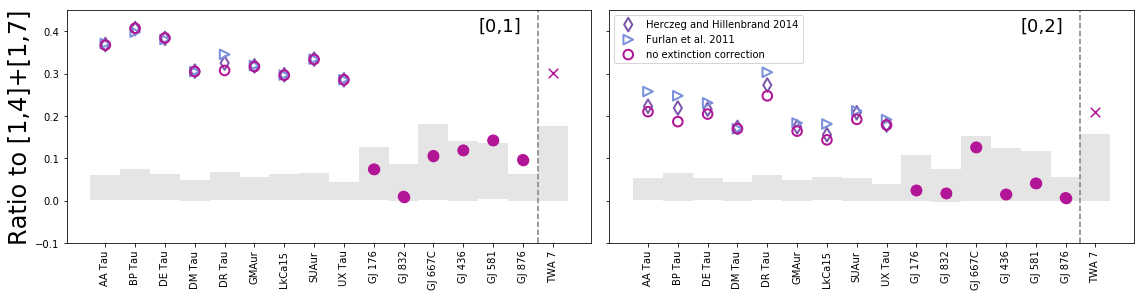}
\caption{The CCF height ratios of [0,1] and [0,2] to that of [1,4]+[1,7] for our selection of T Tauri stars (open symbols), M stars (filled symbols), and TWA 7. The gray areas are the 1$\sigma$ regions for a null result if that progression had no flux and just noise.  They are greater than zero because we select the maximum of the CCF within 15 km s$^{-1}$, which would typically be greater than zero even for random noise.  TWA 7's ratios  matches better with the T Tauri stars, suggesting that its  H$_2$ is also circumstellar.  The ratios for the other progressions are shown in the appendix in Figure \ref{ratios_app}. \label{ratios}}
\end{figure}

 To estimate the 1$\sigma$ limits (the gray areas in Figures \ref{ratios} and \ref{ratios_app}), we used a similar procedure as we used to calculate the significance of detections for TWA 7.  We sampled the noise from each spectrum, calculating a KDE like we did for TWA 7, and created spectra of pure noise to cross-correlate with the template.  We then took the maximum of each CCF within  15 km/s of the RV of that star.  The gray regions represent the inner 68\% of ratios calculated based on those maxima. These 1$\sigma$ regions are biased towards positive numbers, because we chose the maximum CCF value, which even in a normally distributed, random noise sample will bias to positive values.

The [0,1] and [0,2] ratios differentiate the samples most clearly regardless of extinction correction. Based on their data, TWA 7's H$_2$ appears to be more similar to that from the CTTS (Figure \ref{ratios}).  However, other progressions are not as clear.  To analyze all of the ratios, we created a Support Vector Machine  classifier \citep{Platt99probabilisticoutputs} with a 4th order polynomial kernel using the data from the M dwarfs and CTTS --- excluding TWA 7 --- to determine where the H$_2$ was coming from. We then applied this classification scheme to the TWA 7 data.  Based on the observed ratios, this test categorizes TWA 7's H$_2$ as similar to CTTS' 99.2\% of the time for the set uncorrected for extinction, 98.2\% of the time for the set corrected using the extinction values from \citet{HerczegOpticalSpectroscopicStudy2014}, and  98.3\% of the time for the set corrected using the extinction values from \citet{FurlanSpitzerInfraredSpectrograph2011}.  This implies that the TWA 7's H$_2$ is being pumped not only from the core, but also from the wings of the Ly$\alpha$ profile as with CTTS.  We expand upon this in Section \ref{sec_disc}.

\subsection{\texorpdfstring{Estimating the Amount of Circumstellar H$_2$}{Estimating the Amount of Circumstellar H2}}\label{mass_est}

\begin{figure}
\centering
\includegraphics[width=3.8in]{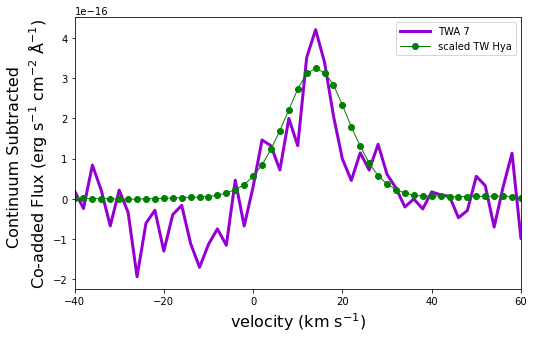}
\caption{The co-added H$_2$ line profile of TWA 7 compared to the co-added line profile of TW Hya, scaled by the best fit ratio of (6.9$\pm$0.7)$\times$10$^{-4}$.\label{coadd}}
\end{figure}

To estimate the amount of warm H$_2$ in TWA 7, we compared its H$_2$ emission with that of a transition disk system, TW Hya.  We chose TW Hya because in comparison with TWA 7, it has a similar age \citep{WebbDiscoverySevenTauri1999}, inclination \citep{PontoppidanSpectroastrometricImagingMolecular2008}, and a relatively similar spectral type \citep{HerczegOpticalSpectroscopicStudy2014}. While we do not think that the line profile of TW Hya would be identical to that of TWA 7, as TW Hya is accreting enough to be measured by conventional methods, it is the best match from the data available.  Our goal was to find a constant scale factor that is the ratio between the H$_2$ line strengths in TWA 7 and those in TW Hya.  We used a least squares fit to calculate this scale factor with the only other free parameter being the radial velocity difference.    We coadded the 19 brightest H$_2$ features from the most prominent progressions --- [1,4], [1,7], [0,1], and [0,2] --- and compared the line flux from the coadded profile to the coadded profile from TW Hya.  We also measured the uncertainty of this ratio by measuring the noise in the spectrum in comparison to the flux.  We found a ratio between the coadded profiles of (6.9$\pm$0.7)$\times$10$^{-4}$.  as shown in Figure \ref{coadd}.   Adjusting for differences in distance, TWA 7 has (2.2$\pm$0.2)$\times$10$^{-4}$ of TW Hya's H$_2$ line strength, and, as a result of its similar inclination and line widths, its H$_2$  luminosity is assumed to be less than TW Hya's by the same factor.  \citet{FranceHubbleSpaceTelescope2012} measure TW Hya's H$_2$  luminosity as (16.2$\pm$2.0)$\times$10$^{29}$ erg s$^{-1}$.  This gives us an H$_2$  luminosity of (3.6$\pm$0.6)$\times$10$^{26}$ erg s$^{-1}$ for TWA 7.  By comparing the flux values measured by \citet{KruczekH2FluorescenceDwarf2017} in star spots to our value for TWA 7' s flux, even if there is a contribution from star spots to this value, we expect that the circumstellar gas is more than 50\% of the total H$_2$ luminosity.

From this scaling factor, we can also put a lower limit on the mass of warm H$_2$ assuming the gas is all circumstellar. The flux observed in a specific H$_2$ line, $F_{obs}$, is a function of the Einstein A value for that emitting transition, $A_{Bl}$, the distance to TWA 7, $d$, the frequency of the emitting transition, $\nu_{Bl}$, and the number of H$_2$ molecules that have been pumped to the required electronic excited state, $N_B$:
\begin{equation}\label{eq_fluor}
    F_{obs} = N_B\frac{A_{Bl}h\nu_{Bl}}{4\pi d^2}
\end{equation}
where $l$ is a lower energy state than $B$.

$N_B$ depends on the number of H$_2$ molecules in the lower state of the pumping transition, $N_X$ and the rate at which those get excited.  This rate is dependent on the oscillator strength $f$, the Ly$\alpha$ flux of TWA 7, and the optical depth of the warm H$_2$. Since we do not know the Ly$\alpha$ flux, the H$_2$ filling factor, or the optical depth, we instead choose to estimate an upper limit for the excitation rate, which turns into a lower limit for $N_X$ as described by Equation \ref{eq_pump}:

\begin{equation}\label{eq_pump}
    %N_{H_2}q(T)=N_X\int\frac{4\pi J_\nu a_\nu}{h\nu_{p}}d\nu =N_XJ_\nu \frac{4\pi^2e^2}{m_ech\nu_{p}}f\leq N_B\sum A_{Bl}
    N_X \frac{4\pi J_\nu}{h\nu_{p}}\frac{\pi e^2}{m_ec}f\leq N_B\sum A_{Bl}
\end{equation}
 where $A_{Bl}$ are all the relevant Einstein A values for the upper state, J$_\nu$ is the Ly$\alpha$ flux at the pumping wavelength, and $\nu_p$ is the frequency of the pumping transition. We do not consider dissociation, as the probability of dissociation for [1,4], [1,7], [0,1], and [0,2] is predicted to be negligible \citep{AbgrallTotaltransitionprobability2000}. This  $N_X$ is  dependent on the total number of warm H$_2$ molecules, $N_{H_2}$, and the temperature, which gets factored into the Boltzmann Equation, $q(T)$:

\begin{equation}\label{eq_part}
    q(T)N_{H_2} =N_X.
\end{equation}
We calculate a $q(T)$   based on the assumption that the H$_2$ is being thermally excited \citep{AdamkovicsFUVIrradiatedDisk2016}. % This is supported by  , who show that the H$_2$ will be heated by layers of  H$_2$O and OH, resulting in a thermally excited layer of H$_2$.

While Ly$\alpha$ flux varies in time, and the HST FUV observation of TWA 7's Ly$\alpha$ flux is contaminated by ISM absorption and geocoronal emission, we estimate that TWA 7's Ly$\alpha$ is less than 0.03 of TW Hya's based on comparison of the spectra at velocities $>$400 km/s \citep{HerczegFarUltravioletSpectraTW2004}.   We estimate the flux observed for a given transition  using our scaling factor found above and the flux observed for TW Hya by \citet{HerczegFarUltravioletSpectrumTW2002}.   All pumping transition properties are described in Table \ref{pros}, while the Einstein A values are from \citep{AbgrallTableLymanBand1993}.   We calculate a separate N$_{H_2}$ for each line flux measured for TW Hya, which then converts into a line flux for TWA 7; we then average these values together to get our final result.  For a gas temperature of 1500 K,   we  get a rough estimate for the minimum amount of warm H$_2$ of $\sim$9.9$\times$10$^{-11}$ M$_\oplus$.  If spread out in a ring with a radius of 0.3 AU --- a radius at which H$_2$ is commonly seen \citep{FranceHubbleSpaceTelescope2012} --- this corresponds to a minimum column density of $\sim$2.8$\times$10$^{15}$ cm$^{-2}$.   This is consistent with the upper limit on H$_2$ column density reported by \citet{InglebyFarUltravioletH2Emission2009} of 3.0$\times$10$^{17}$ cm$^{-2}$ using a less sensitive prism spectrum of TWA 7. Based on the spread of line fluxes, we adopt a range of 10$^{15}$ to 3.0$\times$10$^{17}$ cm$^{-2}$ for the vertical column density of H$_2$ in TWA 7.

\section{Discussion}\label{sec_disc}

The H$_2$ progressions ratios from TWA 7 (Figure \ref{ratios}) more closely resemble that from CTTS than that from M stars. However, these ratios do not guarantee that the H$_2$ is circumstellar.  TWA 7 is much closer in age to the CTTS and is thus likely to have  higher chromospheric activity than an average M star.  Chromospheric activity produces Ly$\alpha$ emission,  which can then excite the H$_2$ in star spots on mid M type stars.  We suggest this is not the primary source of H$_2$ emission in TWA 7 as chromospheric activity affects the core of the Ly$\alpha$ profile significantly more than the wings \citep{LemaireHydrogenLyaLyv2015}.So while there is some Ly$\alpha$ emission in the wings from all of these stars, the amount of flux induced solely from chromospheric activity is likely not enough to excite the outer H$_2$ progressions.  \citet{YoungbloodFUMESIILya2021} looked at how Ly$\alpha$ varied with stellar parameters, showing how increased Ly$\alpha$ is correlated with higher chromospheric activity and lower gravity, both of which are correlated with youth.  However, the profiles from \citet{YoungbloodFUMESIILya2021} show that the ratio of the flux between the peak and the wings can remain constant with varying chromospheric activity and gravity, even if the overall flux changes.    Thus the most likely explanation is that TWA 7 is still weakly accreting circumstellar gas from an inner disk.

Accretion rates for weakly accreting stars are  notoriously hard to measure accurately.  There are cases, like MY Lup, that have FUV accretion signatures but lack optical ones \citep{AlcalaHSTspectrareveal2019}.  Previously, TWA 7 was considered a standard, non-accreting WTTS.  It shows no accretion signatures in the optical. The hot FUV lines seen in TWA 7's HST-COS spectrum, such as C IV or N V, have profiles that do not look like those of CTTS  \citep{ArdilaHotGasLines2013}.  It also lacks the NUV flux and Ca II] $\lambda$2325 emission  of known accreting stars \citep{InglebyNearultravioletExcessSlowly2011,InglebyAccretionRatesTauri2013}. However, most of the accreting gas is expected to be hydrogen in the ground state \citep{MuzerolleDetectionDiskAccretion2000}, so Ly$\alpha$ should be more sensitive to small accretion rates than any other line.  There is a similar system in TWA, TWA 4 B, a K5 star with circumstellar H$_2$ FUV emission discovered by \citet{YangFarultravioletAtlasLowresolution2012} despite not showing obvious accretion signatures.   Given  TWA is close in age to when the typical prototplanetary disk is predicted to evolve into a debris disk, these systems could represent a short-lived phase of disk evolution with residual gas that does not accrete at the high levels detectable in optical spectra.  FUV  spectra of more stars in TWA would allow us to further investigate the gas evolution at this crucial age. %; this is additional evidence that warm gas can exist without accreting at detectable levels

Assuming the H$_2$ we observe is indeed circumstellar, the next question concerns its origin.  One possibility is that the H$_2$ originates from the inward migration, sublimation, and the subsequent photodissociation of H$_2$O ices in comet-like nuclei.  The H$_2$O photodissociates into H, OH, and O, and the newly available H atoms can then reform into H$_2$.  If true, there should also be some  oxygen gas species in the inner disk.  \citet{Riviere-MarichalarGasdustTW2013} give an upper limit on the oxygen mass of 2.3$\times$10$^{-5}$ M$_\oplus$ from Herschel data. This upper limit is more than the oxygen that would accompany the H$_2$ we detect if the H$_2$ originates from dissociated H$_2$O and assuming the warm H$_2$ is confined to the inner few AU of the disk, making this a potentially viable source of the observed H$_2$.  Future observations could better constrain the oxygen mass in the inner disk and allow us to determine whether H$_2$O ice evaporation  is a possible origin of the circumstellar H$_2$ around TWA 7.  Additionally, detection of dust from these comets could lend support to this theory \citep{PearceGastrappinghot2020}.

%If the gas does not originate from the H$_2$O ice evaporation, another possibility is that it is protoplanetary disk gas.

Another possibility is that the H$_2$ we see is residual protoplanetary disk gas.  Regardless of its origin, an H$_2$ formation pathway  is needed to balance ongoing UV photodissociation of H$_2$.  Molecular hydrogen forms most efficiently on grain surfaces,  as in the ISM, but it can also form via gas  phase  reactions (e.g., through H + H$^-$ $\rightarrow$ H$_2$ + e$^-$) when there is less dust surface area available \citep{BrudererSurvivalmoleculargas2013}. 

To explore the possibility of grain surface formation of H$_2$ in the case of TWA 7, we can estimate the upper limit on the surface area of warm grains by looking at its spectral energy distribution (SED).  Although the W3 band from WISE \citep{WrightWidefieldInfraredSurvey2010} shows no excess IR emission from dust \citep{OlofssonResolvingfaintstructures2018, BayoSubmillimetrenoncontaminateddetection2019}, we can put a limit on the amount of warm dust by assuming the dust can generate the equivalent of the 1$\sigma$ uncertainty for the W3 flux.   Under that assumption, we compute the SED using  the model described by \citet{Isellashapeinnerrim2005} to put an upper limit of $\le$5.1$\times$10$^{-8}$ M$_\oplus$ on the amount of  warm ($\sim$1000 K) silicate particles between 1 $\mu$m and 1 mm. We chose a lower particle size limit of 1 $\mu$m, because in more evolved systems, particles smaller than that near the star can get blown away by stellar winds.  Based on this estimate,  grains with radii between 1 $\mu$m and 1 mm could make up a significant surface area, up to a surface area of 10$^{23}$ cm$^2$. If the grains are spread out evenly over the inner 0.3 au, the mass column density is 2$\times$10$^{-5}$  g cm$^{-2}.$  
%, where 1000 K is a typical temperature for small grains within 0.3 au of TWA7. 

With the above constraint on the possible dust content of the inner disk of TWA 7, we can use the results of  \citet{BrudererSurvivalmoleculargas2013} to estimate the H$_2$ reservoir that can be sustained in  the inner disk. In modeling the inner regions of transition disks, \citet{BrudererSurvivalmoleculargas2013} considered two physical models: a dusty inner disk and a very dust-poor inner disk with  dust column densities of $\Sigma_d$=3$\times$10$^{-4}$ g cm$^{-2}$   and 3$\times$10$^{-9}$ g cm$^{-2}$  respectively at 0.3 au. The upper limit on the dust surface density of TWA 7 we find above is an order of magnitude below the surface density of the dusty inner disk model but many orders of magnitude above the surface density of the dust-poor disk model.   Thus the dust-poor disk model provides a relatively conservative estimate of the H$_2$ density allowed for TWA 7.

The other relevant parameter in the \citet{BrudererSurvivalmoleculargas2013} model is the gas surface density.  Figure 6 of \citet{BrudererSurvivalmoleculargas2013} shows the results for a case in which the inner disk and is very dust poor and has a gas column density of 0.3 g cm$^{-2}$  at 0.3\,au.  The H$_2$ fraction in the disk atmosphere is $\sim$3$\times$10$^{-6}$  relative to hydrogen or an  H$_2$ column density of N$_{\rm H_2}$=3$\times$10$^{18}$ g cm$^{-2}$.    \citet{BrudererSurvivalmoleculargas2013} do not show the temperature of the H$_2$, although much of it is likely to be warm, as the disk is dust poor, and dust is a coolant for the gas through gas-grain collisions. If 0.1\% of the total H$_2$ column is warm ($\sim$1500 K),   this scenario predicts a warm H$_2$ mass similar to that inferred for TWA 7. 

Note that this result is obtained despite using a model with a dust density several orders of magnitude below our dust upper limit. Thus, it seems plausible that even a dust-poor inner disk can sustain a warm H$_2$ column density in the range we estimate for TWA 7 in Section \ref{mass_est}. While  the models from \citet{BrudererSurvivalmoleculargas2013} were not tuned specifically to TWA 7's parameters --- the model assumes  a hotter 10 L$_\odot$ star and a given polycyclic aromatic hydrocarbons (PAH) abundance --- these two factors should impact the H$_2$ production in opposite ways: the higher UV flux of the more massive star enhances photodestruction of H$_2$, while the PAH abundance enhances H$_2$ production.  We therefore believe it is plausible that the H$_2$ we detect is  sustained via some combination of gas phase reactions in the circumstellar environment of  TWA 7.   Future observations between 3 and 12 microns with telescopes like JWST could detect PAHs in the disk and lend further support to this possibility  \citep{SeokPolycyclicAromaticHydrocarbons2017}.

Although we do not have the requisite measurements to conclusively determine why there is is warm H$_2$ in the circumstellar environment of TWA 7, 
regardless of its origin, warm gas in a region without detectable warm dust is not unique to this star.  Primordial warm H$_2$ is detected inside the inner edge of the dust disk  in transitional disk systems \citep{FranceHubbleSpaceTelescope2012, ArulananthamUVtoNIRStudyMolecular2018}.  Warm CO has also been detected in these regions \citep{PontoppidanSpectroastrometricImagingMolecular2008, SalykCORovibrationalEmission2011}.  Clearly, warm gas can outlast detectable amounts of warm dust.     Thus, the physics resulting in warm gas in the cavities of transitional disks  could also be the cause  of the H$_2$ we detect in TWA 7.

\section{Conclusions}
We have detected molecular hydrogen from four progressions ([1,4], [1,7], [0,1], and [0,2]) in TWA 7, a known debris disk system.  The ratios between CCF peaks of the detected H$_2$ progressions (Figure \ref{ratios}) resemble those from CTTS.  This suggests that the H$_2$ in TWA 7 is circumstellar, as it is for CTTS. This is highly unexpected, because H$_2$ is not typically detected in debris disk systems.  This star joins a small group of systems that have H$_2$ but are not accreting by typical diagnostic standards.   Assuming the H$_2$ is circumstellar, we have estimated a column density of 10$^{15}$ to 3.0$\times$10$^{17}$ cm$^{-2}$.  While we cannot determine the origin of the gas conclusively, it is likely to be generated from residual protoplanetary disk gas.

\acknowledgements
Based on observations with the NASA/ESA Hubble Space Telescope obtained at the Space Telescope Science Institute, which is operated by the Association of Universities for Research in Astronomy, Incorporated, under NASA contract NAS5-26555. Support for program number (GO-15310) was provided through a grant from the STScI under NASA contract NAS5-26555. GJH is supported by by general grant 11773002 awarded by the National Science Foundation of China.  L.F. would like to thank Andrea Isella for help with the S.E.D. modeling.  This research has made use of the VizieR catalogue access tool, CDS, Strasbourg, France. The original description of the VizieR service was published by  \citep{WengerSIMBADastronomicaldatabase2000}.  This research has made use of the SIMBAD database,operated at CDS, Strasbourg, France. This research has made use of the NASA/ IPAC Infrared Science Archive, which is operated by the Jet Propulsion Laboratory, California Institute of Technology, under contract with the National Aeronautics and Space Administration. This work has made use of data from the European Space Agency (ESA) mission {\it Gaia} (\url{https://www.cosmos.esa.int/gaia}), processed by the {\it Gaia} Data Processing and Analysis Consortium (DPAC, \url{https://www.cosmos.esa.int/web/gaia/dpac/consortium}). Funding for the DPAC has been provided by national institutions, in particular the institutions participating in the {\it Gaia} Multilateral Agreement. 

%\citet{MatraUbiquityStellarLuminosity2019} conclude that the CO in the disk, based on its total mass, is released by collisions of planetesimals; however, this conclusion assumes no shielding from gas in the inner disk.  

\facility{HST (COS, STIS)}

\software{SpecTres \citep{CarnallSpectResFastSpectral2017}, NumPy \citep{oliphant2006guide, van2011numpy}, Scikit-learn \citep{scikit-learn}, Pandas \citep{reback2020pandas}, Scipy \citep{VirtanenSciPyFundamentalAlgorithms2019}, Matplotlib \citep{Hunter:2007} }

\restartappendixnumbering

\appendix

\section{CCF Ratios}
\begin{figure}
\centering
\includegraphics[width=6.4in]{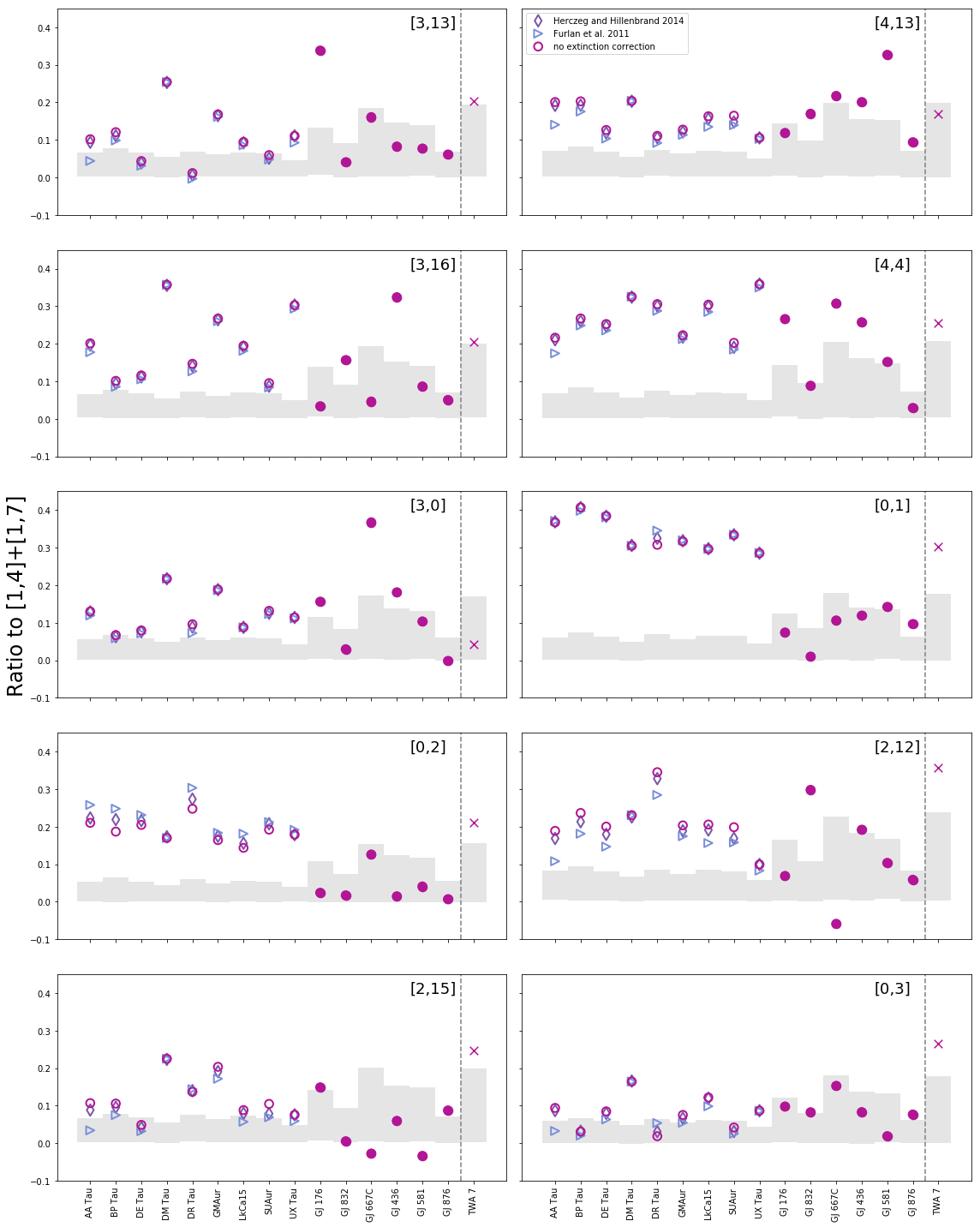}
\caption{The CCF height ratios of non-central progressions to that of [1,4]+[1,7] for our selection of T Tauri stars (open symbols), M stars (filled symbols), and TWA 7. The gray areas are the 1$\sigma$ regions for a null result if that progression had no flux and just noise.  They are greater than zero because we select the maximum of the CCF within 15 km s$^{-1}$, which would typically be greater than zero even for random noise.  TWA 7's ratios  matches better with the T Tauri stars, suggesting that its  H$_2$ is also circumstellar. \label{ratios_app}}
\end{figure}

We took the ratio of the CCF maximum of all the non-central progressions
 to the ratio of the CCF maximum of [1,4]+[1,7] for all the stars with previously detected H$_2$ in our sample.  In Figure \ref{ratios_app}, we plot all of these ratios.  TWA 7's ratios were statistically more similar to that of the CTTS than that of the main sequence M dwarfs.  We describe this analysis in detail in Section \ref{origin}.
 
 [0,1] and [0,2] have the most easily detectable flux because of a combination of several factors regarding the pumping transition shown in Table \ref{pros}: relatively close to the center of Ly$\alpha$, high oscillators strengths, and low energy levels for the lower state of the ground pumping transition.  
 
\bibliography{h2indisks_url}
\end{document}